# Regulation of interferon production as a potential strategy for COVID-19 treatment

Xiaobing Deng[a#*], Xiaoyu Yu[b#], Jianfeng Pei[c*]

[a] Infinity Intelligent Pharma Co., Ltd. Beijing, 100083, China. dengxb@iipharma.cn.
[b] Peking-Tsinghua Center for Life Sciences, Peking University, Beijing 100871, China.
[c] Center for Quantitative Biology, AAIS, Peking University, Beijing, 100871, China. jfpei@pku.edu.cn.
[#] XB Deng and XY Yu contributed equally to this work.

**Abstract:** Regulating the upstream of the cytokines production could be a promising strategy to the treatment of COVID-19. We suggest to pay more attention to the dysregulated IFN-I production in COVID-19 and to considerate cGAS, ALK and STING as potential therapeutic targets preventing cytokine storm. Approved drugs like suramin and ALK inhibitors are worthy of clinical trials.

**Introduction and Reasonable Inference.**

Novel Coronavirus SARS-CoV-2 is a single strand RNA virus responsible for the ongoing severe respiratory illness and pneumonia-like infection in human worldwide, which has been regarded as a global health emergency (*1*). SARS-CoV-2 belongs to the β-coronavirus cluster, same as SARS-CoV and MERS-CoV (*2*). Type-I interferon (IFN-I) activates intracellular pathogen defense and influence the development of innate immunity and adaptive immunity. The DNA sensor cyclic GMP–AMP synthase (cGAS) and its downstream effector STING (stimulator of interferon genes, also known as ERIS/ MITA) control transcription of many inflammatory mediators, including type I and type III interferons (*3, 4*). However, abnormal recognition or interpretation of danger-associated molecular patterns (DAMPs) by cGAS, like self-DNA released by apoptotic cells, also arises human inflammatory disease (*5*). And STING gain-of-function mutants is responsible for various devastating autoimmune diseases known as STING-associated vasculopathy with onset in infancy (SAVI)(*6-8*). Excessive inflammation induced by STING disorder damages the body's own cells and tissues, causing SAVI phenotype including systemic inflammation, destructive skin lesions, and interstitial lung disease (*6*). In SARS infected mice model, rapid spreading virus ultimately induced a delayed IFN-I production and promoted severe disease in the late state by boosting accumulation of pathogenic monocyte-macrophages, resulting in lung immunopathology, vascular leakage, and suboptimal T cell responses(*9*). In SARS patients, high IFN and IFN-stimulated chemokine levels, plus robust antiviral IFN-stimulated gene expression, accompanied early SARS sequelae and poor clinical course (*10*). In Sepsis, a life-threatening organ dysfunction caused by infection, and silica-induced lung inflammation, STING dependent disordered IFN-I response was also reported (*11, 12*). In COVID-19, the lung damage level and the level of cytokines in both ICU patients and non-ICU patients were higher than those in healthy adults(*2*). Similar phenomena were also found in SARS patients and silica-induced lung inflammation patients.

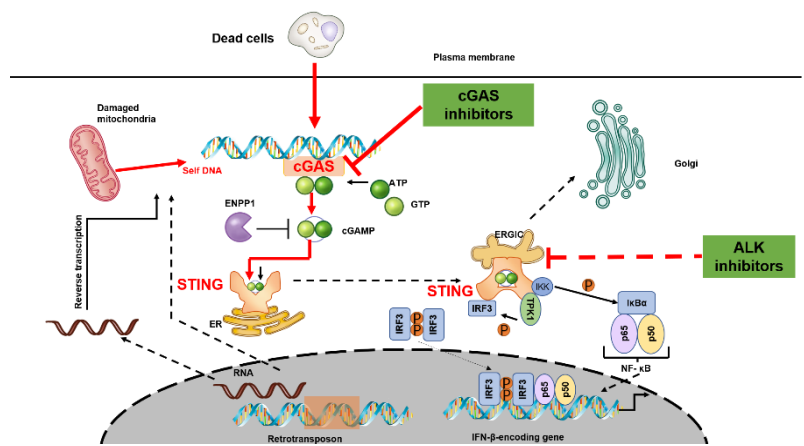

Figure 1. The proposed cGAS-STING-IFN production pathway related to COVID-19. Modified from ref (*13*).

Given that preventing cGAS-STING pathway from aberrant activation may be a suitable strategy for treatment of severe lung diseases induced by SARS-CoV, SARS-CoV-2 or other pathogens, we suppose that several proteins in the pathway are potential targets for COVID-19 (Figure 1). STING is a popular drug target as immunosuppressants in recent years(*14, 15*). Several STING regulatory molecules were developed by academy and industry(*16*). Unfortunately, none STING directly-target molecules has been marketed yet.

**Results and Discussion**

To obtain potential STING safety inhibitors, we screened the FDA-approved drug library using virtual screening. We selected 6 compounds through the screening.  The molecular structures of these 6 compounds are shown in Figure 2.

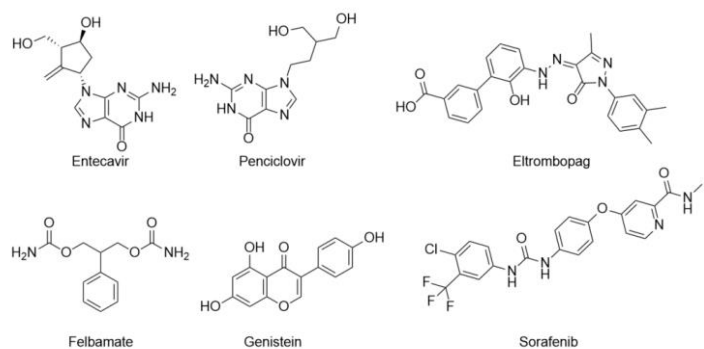

Figure 2. potential STING inhibitors supposed by virtual screening.

The selected compounds were evaluated with two cell-based assays.  Sorafenib potently inhibits vaccinia virus induced IFN-β production in THP1 cells (Fig 3A).  And in Hela cells, sorafenib inhibited dsDNA induced IRF3 phosphorylation dose dependent after incubating for 12 hours or 20 hours (Fig 3B). These data indicated that, sorafenib is a potent inhibitor of STING pathway.

Further direct STING-Sorafenib binding assays might give more evidence.

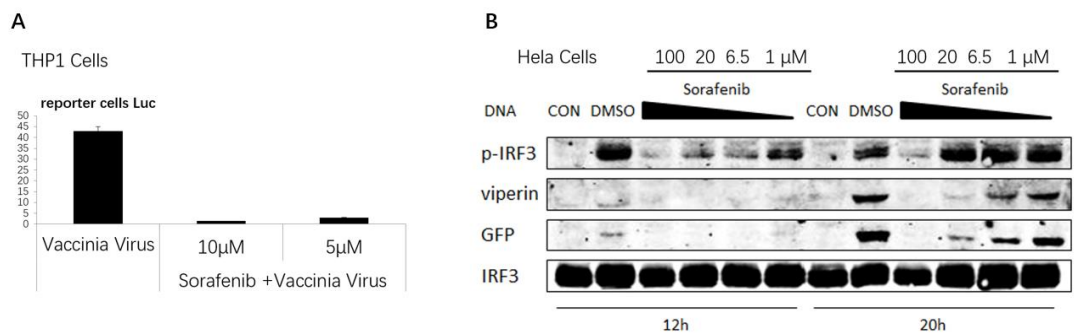

Figure 3. Sorafenib is shown to be a potent STING inhibitor

Using drugs targeting STING regulatory proteins to regulate IFN-I production is also a potential strategy to meet an urgent need for marketed drugs. Recently, anaplastic lymphoma kinase (ALK) inhibitors were reported to be a specific and effective STING antagonists both *in vitro* and *in vivo* (*17*). ALK directly interacts with epidermal growth factor receptor (EGFR) to trigger serine-threonine protein kinase AKT phosphorylation and activates interferon regulatory factor 3 (IRF3) and nuclear factor κB (NF-κB) signaling pathways, enabling STING-dependent rigorous inflammatory responses (*12*). Thus, ALK-targeted drugs have great potential in treating moderate to severe lung inflammation in COVID-19.

cGAS, which is the cytosolic DNA sensor upstream STING, is also a potential therapeutic target for IFN-I regulating. Suramin, an approved antiparasitic drug, was reported to be an effective cGAS antagonist by displacing the bound DNA from cGAS (*18*). Suramin was reported as an entry inhibitor for various of viruses including DNA virus and RNA virus(*19-23*). Clinical trial application of Suramin had been submitted in China on 21th Feb(*24*).

Cytokines directed antagonists, such as Adalimumab (TNF-α) and CMAB806 (IL-6) have been applied in clinical trials against COVID-19(*24*). Diverse cytokines are involved in the pathology of

COVID-19. Thus, regulating the upstream of the cytokines production could be a promising strategy. Collectively, we suggest to pay more attention to the dysregulated IFN-I production in COVID-19 and to considerate cGAS, ALK and STING as potential therapeutic targets preventing cytokine storm. Approved drugs like suramin and ALK inhibitors are worthy of clinical trials.

**Experimental section.**

**Virtual Screening.** The c-di-GMP bond to STING crystal complex was used as receptor (PDB code:4EMT). And the commercial FDA-approved drug data containing 2684 compounds was downloaded from Selleck website. A two-step molecular docking scheme was performed, which has been successfully used in discovering active molecules in other systems(*25*). Virtual screening using FDA-approved drugs against cGAS and ALK were also performed but no potential compounds other than known inhibitors were found. The general procedure was as following: all compounds were docked into the ligand binding pockets with the rigid-body docking approach of the DOCK 6.1(*26*) program using the default parameters. Two hundred compounds with the lowest estimated Ki values were selected for the next step. The AutoDock step of virtual screening was performed with the Lamarckian genetic algorithm using the following parameters: number of genetic algorithm runs, 20; number of individuals in population, 150; maximum number of energy evaluations, 1 750 000; maximum of generations, 27 000. Size of grid box was 60 × 60 × 60 (number of grid points in *xyz*-coordinates). Center of grid box was at the center of ligand binding site. The first 20 compounds with the lowest estimated Ki values were selected. The binding conformations of these compounds were exported and manually evaluated according to the following criteria: (1) The compound formed at least two hydrogen bonds. (2) The compound was not a polypeptide.

**Cell-Based Assay:**

**Type I-IFN bioassay.** Type I-IFN concentration was measured as previously described(*27*). THP1 cells were starved by 2% FBS containing DMEM for 48 h before infected with vaccinia virus. reporter cells were planted to 96-well plates and incubated with human cell culture supernatants. Recombinant human IFN-β (R&D Systems) was used as standards. 4 h later, cells were lysed and measured by Luciferase Reporter Assay System (Promega).

**Western blot.** HeLa cells planted on 10-cm dishes (1×107 cells/dish) were transfected with the indicated expression plasmids for 20 h. HeLa cells were treated with dsDNA and Sorafenib for the indicated times. Cells then were lysed in lysis buffer (1% Triton X-100, 150 mM NaCl, 12.5 mM β-glycerolphosphate, 1.5 mM MgCl2, 2 mM EGTA, 10 mM NaF, 1 mM Na3VO4, 2 mM DTT) containing protease inhibitors. Lysates were centrifuged and the supernatants were incubated with anti-Flag beads or with Protein A/G Sepharose (Amersham) plus anti-IRF3, anti-p-IRF3, anti-viperin and anti-GFP antibodies for at least 4 h. The beads were washed with cold PBS for three times and eluted with DTT-containing SDS sample buffer by boiling for 10 min before Western blot analysis.

**Acknowledgements:**

The literature inference in this work was supported by an AI platform named 'Drug Discovery Brain' of Infinity Intelligent Pharma Co., Ltd. We thank Dr. Youjun Xu and Dr. Weilin Zhang for their help in computer programming and data processing.